\documentclass[aps,prd,preprint,superscriptaddress]{revtex4}
\usepackage{feynmf}
\usepackage{amssymb}
\usepackage[dvips]{graphicx}
\newcommand{\be}{\begin{equation}}
\newcommand{\ee}{\end{equation}}
\newcommand{\B}{{\cal B}_{\xi}}

\begin{document}
\begin{fmffile}{pippo}

\preprint{TRINLAT-02/07}
\preprint{UPRF-2003-01}

\title{Gauge Theories on a 2+2 Anisotropic Lattice}

\author{Giuseppe Burgio}
\affiliation{School of Mathematics, Trinity College, Dublin 2, Ireland.}

\author{Alessandra Feo}
\affiliation{School of Mathematics, Trinity College, Dublin 2, Ireland.}
\affiliation{Dipartimento di Fisica, Universit\`a di Parma and INFN Gruppo
             Collegato di Parma, Parco Area delle Scienze, 7/A, 43100 Parma,
             Italy}

\author{Mike Peardon}
\author{Sin\'ead M. Ryan}
\affiliation{School of Mathematics, Trinity College, Dublin 2, Ireland.}
\collaboration{The TrinLat Collaboration}
\noaffiliation

\date{\today}

\begin{abstract}
The implementation of gauge theories on a four-dimensional anisotropic lattice with two distinct lattice spacings is
discussed, with special attention to the case where two axes are finely and two axes are coarsely discretized. 
Feynman rules for the Wilson gauge action are derived and the renormalizability of the theory and the recovery of the continuum
limit are analyzed. 
The calculation of the gluon propagator and the restoration of Lorentz invariance in on-shell states is presented to one-loop
order in lattice perturbation theory for $SU(N_c)$ on both 2+2 and 3+1 lattices.
\end{abstract}

\pacs{}
\maketitle

\section{Introduction}
\label{sec:intro}
The anisotropic lattice is a popular tool in numerical simulations. The usual approach is to make the 
temporal lattice spacing fine while keeping the spatial directions relatively coarse. 
The extra temporal resolution in this 3+1 discretization scheme is useful when measuring two- and 
three-point correlation functions of particles which decay rapidly.
This is particularly useful for heavy hadrons and glueball states, for which the signal to noise ratio 
decreases rapidly with time. 
The glueball spectrum~\cite{Morningstar:1999rf,Morningstar:1997ff} was an
early success for this approach and, more recently, heavy quark systems
have been studied using anisotropic lattices~\cite{Fingberg:1998qd,Liao:2001yh,Juge:1998nd,Juge:1999ie,
Juge:1999ar,Drummond:1998fd,Manke:1998qc,Manke:1999ru,Drummond:1999db,Collins:2001pe,Harada:2001ei,
Hashimoto:2003fs}.

In this paper we consider a generalization of the anisotropic approach to include lattices with a 
2+2 discretization. The temporal and one spatial direction are made fine, keeping the remaining two 
spatial directions coarse. The motivation for this choice is to explore the feasibility of calculating 
decays which produce high-momentum daughter particles. These include the phenomenologically interesting 
transistions $B\rightarrow\pi l\nu$ and $B\rightarrow K^\ast\gamma$. CKM matrix elements are determined from 
such exclusive decay processes through a combination of experimentally 
measured branching fractions and theoretical calculations of form factors. The light daughter hadrons in these 
decays have non-zero momentum and so in a lattice calculation of the nonperturbative form factors 
there are cutoff effects proportional to this momentum in units of the lattice spacing. 
Therefore $B\rightarrow\pi l\nu$ has discretisation 
errors proportional to $ap_\pi$ where $a$ is the lattice spacing. 
However, the range of momenta reliably reached by current experimental and lattice data do not overlap. 
Lattice calculations work best with $p_\pi\leq 1$GeV but the bulk of 
experimental data for the exclusive decay, $B\rightarrow\pi l\nu$ lie at $p_\pi\geq 1.5$GeV. 
To avoid model-dependence kinematic cuts can be applied, restricting the range of lattice data 
to momenta where the calculation is reliable~\cite{El-Khadra:2001rv}, awaiting improved 
experimental results.
Alternatively, the lattice data can be extrapolated to match 
experiment~\cite{Bowler:1999xn,Abada:2000ty,Aoki:2001rd} but 
this introduces model-dependence and increases substantially the 
systematic error in lattice calculations and therefore in $|V_{ub}|$. 
For $B\rightarrow K^\ast\gamma$ 
the situation is more acute since this decay happens at the maximum recoil momentum of the 
daughter meson, far from the range accessible to current lattice calculations. Therefore it is necessary to 
extrapolate well outside the range of reliable data~\cite{Burford:1995fc,Becirevic:2002zp}, once again 
introducing model-dependence and increasing the systematic errors. For this reason it has not been 
widely studied using lattice methods and calculations to date have used isotropic lattices. In both 
cases the difficulty for lattice 
calculations is that errors proportional to the momentum of the order $ap$, grow 
quickly. These must be controlled to make phenomenologically relevant lattice calculations. 
The advantages then of the 2+2 lattice are twofold. Firstly, the fine temporal lattice spacing serves the 
same purpose as in the 3+1 case: correlation functions should be accurately determined while keeping 
computational costs modest. Secondly, making one spatial direction fine and injecting all momentum 
along that direction keeps discretization errors of ${\cal O}(ap)$ small for high momenta.  

The transition $B\rightarrow\pi l\nu$ has been calculated using a 3+1 
discretisation scheme and the improved resolution in the temporal direction led to higher-momentum 
($0\leq p_\pi\leq 1.5$GeV) particles being reliably simulated~\cite{Shigemitsu:2002wh}. 
Both this calculation and the isotropic calculations which were reviewed in Ref.~\onlinecite{Ryan:2001ej} find that 
one of the 
largest systematic errors in the range of accessible momenta is due to the chiral extrapolation. The 
2+2 discretization does not address this issue but it is hoped that it will further extend the range of 
momentum available to lattice calculations and that statistical precision will be enhanced. 

The paper is organized as follows. In Section~\ref{sec:aniso} a general anisotropic formulation for 
$SU(N_c)$ Yang-Mills theory in four dimensions is presented. 
The differences between 3+1 and 2+2 discretisations are 
discussed in terms of the lattice symmetries and the parameter tuning required. 
Section~\ref{sec:pert} contains the framework of the perturbative calculation, already outlined 
in Ref.~\onlinecite{Burgio:2002uq}. Although the goal is a 2+2 discretization, the 
Feynman rules and the analytic procedure given are completely
general and allow an exact treatment of the calculations. 
In Section~\ref{sec:results} we present our results and include a comparison 
with other work. The challenging algebraic manipulations were performed on a computer 
using a symbolic code which handles the dependence on the anisotropy analytically. 
The core of the code is similar to that used in Ref.~\onlinecite{Alles:1994yh}, 
while the treatment of the lattice 
integrals follows from Ref.~\onlinecite{Burgio:1996ji}.
Section~\ref{sec:conclude} contains our conclusions, while some technical
details are given in Appendices~\ref{app:4glver} and~\ref{app:Integrals}.

\section{Yang-Mills theories on a 2+2 anisotropic lattice}
\label{sec:aniso}
In this section, formulations of $SU(N_c)$ Yang-Mills theory on general
orthogonal lattice types are considered. A generalisation of the Wilson action 
for $SU(N_c)$ Yang-Mills theory on the lattice is given by
\begin{equation}
	S_{\rm W} = \beta\sum_{n,\mu\nu}c_{\mu\nu}\left
           (1-\frac{1}{2N_c}\mbox{Tr}(P_{\mu\nu}(x)+P^\dagger_{\mu\nu}(x))\right),
	\label{eqn:S-Wilson}
\end{equation}
where $\beta = 2N_c/g^2$ with $g^2$, the lattice coupling constant and $N_c$ the number of 
colors. $P_{\mu\nu}$ is the plaquette in the $(\mu,\nu)$ plane;
\begin{equation}
	P_{\mu\nu}(x) = U_\mu(x)U_\nu(x+\hat{\mu})
			U^\dagger_\mu(x+\hat{\nu})U^\dagger_\nu(x). 
	\label{eqn:plaquette}
\end{equation}
The six coefficients, $c_{\mu\nu}$ in Eq.~(\ref{eqn:S-Wilson}) will describe the 
anisotropy class. 

If some sub-sets of these
parameters are given identical values, the lattice action may have symmetries
under the interchange of axes.  In particular, if the two identities
\be
  c_{12}=c_{23}=c_{31} (=c_{cc}) \mbox{\hspace{3em}and\hspace{3em}} 
  c_{41}=c_{42}=c_{43} (=c_{cf})
  \label{eqn:3+1}
\ee
are imposed, the lattice
action is symmetric under the cubic point group, and the 3+1 anisotropy class
is realised. Similarly, if the identity 
\be
c_{13}=c_{14}=c_{23}=c_{24} (=c_{cf})
  \label{eqn:2+2}
\ee
is
imposed, the lattice action transforms trivially under elements of the group
$C_{4\nu}\otimes C_{4\nu}$, with the first constituent group comprising the
rotations and reflections in the (1,2) plane and the second being those in the
(3,4) plane. This anisotropy class is denoted 2+2. 

For phenomenologically motivated reasons, as discussed in
Section~\ref{sec:intro}, consideration will be restricted to these cases
where only two distinct lattice spacings are permitted. The coarse and fine
lattice spacing will be denoted $a_c$ and $a_f$ respectively. 
In this paper the overall scale $a$ is chosen to be $a_f$. 
The implementation of a 3+1
anisotropy, in which the lattice spacing in the temporal direction is made fine
keeping the spatial lattice spacing coarse, has been widely discussed in the
literature~\cite{Karsch:1982ve,GarciaPerez:1997ft}.

In the continuum, the corresponding sub-groups of $O(4)$, the rotations in
Euclidean space that are generated by demanding the same symmetries in the
coefficients of the continuum dimension-four operators, are $O(3)$ for the
class described in Eq.~(\ref{eqn:3+1}) and $O(2)\otimes O(2)$ for Eq.~(\ref{eqn:2+2}). 
If the operators which transform trivially under these
symmetries are enumerated there is one operator for the Euclidean group, (Tr
$F_{\mu\nu}F_{\mu\nu}$), two operators for the $O(3)$ spatial rotation group
(most conveniently denoted Tr $E_i E_i$ and Tr $B_i B_i$ for chromoelectric and
chromomagnetic) and three operators for the $O(2)\otimes O(2)$ case (Tr
$F_{f\!f'}F_{f\!f'}$, Tr $F_{fc}F_{fc}$ and Tr $F_{cc'}F_{cc'}$). 

The recovery of Euclidean invariance in the low-energy physics of an 
anisotropic lattice requires a parameter tuning, in contrast to the isotropic 
case. Additional differences arise between the 3+1 lattices, for which a
one-parameter tuning suffices and 2+2 lattices which necessitate a
two-parameter tuning. The importance of parameter tuning, in particular for the
2+2 case is emphasised in subsection~\ref{subsec:tuning}. 

Finally, it is interesting to note that taking the anisotropy $\xi\to \infty$ 
yields, up to a gauge transformation, 
the Hamiltonian limit of the theory on a 3+1 lattice. For the 2+2 anisotropy 
class, this limit can not be taken naively; a theory in which only two of the 
four dimensions are discretised is not regularised. 

\subsection{2+2: the need for tuning}
\label{subsec:tuning}
As the continuum limit is approached for the 3+1 lattice theory, the physics of
the Euclidean invariant Yang-Mills theory is reproduced, provided care is taken
to account for the different grid spacings, $a_c$ and $a$. 
This can be achieved by tuning the relative weights of the two sets of
coefficients in Eq.~(\ref{eqn:3+1}) to ensure the ratio of scales, 
$\xi = a_c/a$
measured by a physical probe takes its desired value or alternatively,
an arbitrary choice of the two parameters $c_{cc}$ and $c_{cf}$ can be taken
and the ratio of scales measured {\it post-hoc}. The 3+1 anisotropic theory is
certain to be in the same universality class as the desired theory since the
single free parameter (the relative weight of $c_{cc}$ and $c_{cf}$) determines 
the ratio of scales $\xi$. 

For the 2+2 case,
an important distinction arises; while there are only two distinct
dimension-four continuum operators in the 3+1 class, there are three for the 2+2
case. As a result, the recovery of a Euclidean invariant continuum theory is 
not guaranteed since there are two free parameters and only a single ratio of 
scales. As a result, the general 2+2 lattice theories can lie in a larger
universality class than the continuum four-dimensional Yang-Mills theory. 

For this reason, care must be taken to ensure the recovery of Euclidean 
invariance in the continuum limit for a 2+2 simulation. The relative weights
of the three operators in the action must be determined to ensure Lorentz
invariance in on-shell Green's functions. 
This tuning can be achieved perturbatively or by restoring symmetries in a non-perturbative
calculation of, for example the static inter-quark potential.
In the remainder of this paper, perturbation theory is used to determine the
parameters in the action. A paper describing the nonperturbative tuning 
of the parameters is in preparation~\cite{Burgio:2002aa}. 
\subsection{$c_{\mu\nu}$ from perturbation theory}
\label{subsec:def_cmunu}
At tree level, it is straightforward to determine the values of the
coefficients in the action. For the 3+1 lattice action, they are 
$c_{cc}=1/\xi$, $c_{cf}=\xi$ and for the 2+2 case, they are
$c_{cc}=1/\xi^2$, $c_{cf}=1$ and $c_{f\!f}=\xi^2$. 
The one-loop definitions of $c_{\mu\nu}$ can then be parameterised as
\begin{equation}
	c_{\mu\nu}^{(2+2)} = \left\{\begin{array}{llr}
	     \xi^2 & (1+\eta^{(1)}_{f\!f}g^2 + O(g^4)) & \mu , \nu \mbox{ fine} \\
	     1     & (1+\eta^{(1)}_{cf}g^2 + O(g^4)) & \hspace{3em}\mu\mbox{ coarse(fine)}, 
						   \nu\mbox{ fine(coarse)}\\ 
   \frac{1}{\xi^2} & (1+\eta^{(1)}_{cc}g^2 + O(g^4)) & \mu , \nu \mbox{ coarse}
			    \end{array}\right.
\label{cs}
\end{equation} 
and 
\begin{equation}
	c_{\mu\nu}^{(3+1)} = \left\{\begin{array}{llr}
	     \xi & (1+\eta^{(1)}_{cf}g^2 + O(g^4)) & \hspace{3em}\mu\mbox{ coarse(fine)}, 
						   \nu\mbox{ fine(coarse)}\\ 
      \frac{1}{\xi} &(1+\eta^{(1)}_{cc}g^2 + O(g^4)) & \mu , \nu \mbox{ coarse}
			    \end{array}\right.
\label{cs1}
\end{equation}
In both cases, an overall multiplicative weight can be absorbed into a
redefinition of the field integration variables, and this is redundant in the
action-tuning procedure since a symmetry is being enforced in on-shell Green's 
functions. This change of variables will become important once matrix-element 
matching of gluon fields is being performed. 

\section{Perturbation theory}
\label{sec:pert}
We can now proceed to discuss Feynman rules for a general four-dimensional
anisotropic Wilson-like action, already outlined in Ref.~\onlinecite{Burgio:2002uq}.
We follow the notation given in Ref.~\onlinecite{Book:Rothe} and
the interpretation of the anisotropy as a difference in momentum cut-offs will be our guideline.
The matching of the lattice gluon action in Eq.~(\ref{eqn:S-Wilson}) with its continuum counterpart is made
clearer by expressing the link variables $U_\mu(n) = e^{i \phi_\mu(n)}$ in terms of dimensionless fields
\be
	\phi_\mu(n) = \phi^b_\mu(n) T^b \, ,
	\label{phield}
\ee 
where $T^b$ are the $SU(N_c)$ generators in the fundamental representation satisfying the relations,
$\displaystyle{[T^a,T^b] = if^{abc}T^c}$ and $\displaystyle{\rm Tr}(T^aT^b)=\frac{1}{2}\delta^{a b}$. The 
dimensionful gluon fields, $A_\mu(x)$ can then be reintroduced in Eq.~(\ref{phield}) 
using the relation
\be
	\phi^b_\mu(n) = g\, a\, \xi_{\mu}\, A^b_{\mu}(x) \, ,
\ee
where
\be 
	\xi_{\mu}= \left\{ \begin{array}{lr}
        	      1   & \mbox{$\mu$ fine } \\
	              \xi & \hspace{3em}\mbox{$\mu$ coarse} \end{array} \right.
	\label{asind}
\ee
is the anisotropy index. By taking into account the Jacobian 
\be
	\sum_n = \frac{1}{\xi^d a^4} \sum_x \, ,
\ee 
where $d$ denotes the number of coarse directions,
one easily gets Eqs.~(\ref{cs}) and~(\ref{cs1}).
The continuum limit of the Eq.~(\ref{eqn:S-Wilson}) is then written as 
\be
	S_{\rm W} = \sum_n \frac{1}{4} c_{\mu\nu} \hat{F}^b_{\mu\nu}(n)\hat{F}^b_{\mu\nu}(n) + 
			{\cal O} = \frac{1}{4} \sum_{x} F^b_{\mu\nu}(x) F^b_{\mu\nu}(x) + O(a^2) \,  ,
	\label{gluon}
\ee
where $\cal{O}$ is an irrelevant operator in the continuum involving terms with three or more 
gluon fields from which the non-trivial contribution to the Feynman rules will arise, while
\be 
\hat{F}^b_{\mu\nu}(n) = \hat{\partial}^R_{\mu} \phi^b_{\nu}(n) - \hat{\partial}^R_{\nu} 
\phi^b_{\mu}(n) -g f^{bcd} \phi^c_{\mu}(n) \phi^d_{\nu}(n) \, , 
\label{eq1}
\ee
and 
\be
	\hat{\partial}^{R}_{\mu} \varphi(n) = \varphi(n+\hat{\mu}) - \varphi(n)  
\ee 
are the dimensionless field strength and lattice right derivative, whereas
\be
	F^b_{\mu\nu}(x) = \partial^R_{\mu} A^b_{\nu}(x) - \partial^R_{\nu} 
			   A^b_{\mu}(x)- g f^{bcd} A^c_{\mu}(x) A^d_{\nu}(x) ,
\ee
and 
\be 
\partial^{R}_{\mu} \varphi(x) = \frac{1}{a\, \xi_{\mu}} \left(\varphi(x + a\, \xi_{\mu}\, \hat{\mu}) - 
\varphi(x)\right) 
\ee 
are their dimensionful equivalents. 

As is well known, the Wilson action has $O(a^2)$ discretization errors in the 
evaluation of various physical quantities. In addition, the lattice regularization gives rise to
finite renormalization coefficients when compared to other continuum schemes. 
These effects can be reduced by adding irrelevant 
operators in the action which reduce discretization artefacts,
or by improving the convergence of renormalization coefficients to better match continuum 
quantities~\cite{Symanzik:1983dc,Symanzik:1983gh}. 
In any of these cases an exact
and  completely algebraic treatement of the Feynman rules is always viable,
even at orders higher than one~\cite{Luscher:1995np,Alles:1997cy,Alles:1998is,Christou:1998ws,Follana:2000mn}. As for the lattice 
integrals, while a completely numerical evaluation presents no problem at first order, 
any one-plaquette action will change the structure of the 
$O(a^2)$ (irrelevant) terms in 
the four and higher gluon vertices 
but will not change the propagators, the three-gluon vertex, 
the measure or the gauge-fixing terms,
which are all fixed by the Haar measure and the naive matching with the continuum 
limit~\cite{Ellis:1984af}.
This means that once the first-order analytic technique for the Wilson 
action is developed it can be applied to any other one-plaquette action. 
For gauge actions with Wilson loops which extend two grid spacings, the technique given in 
Ref.~\onlinecite{Burgio:1996ji} (for fermion propagators) can be 
adapted while for actions with even larger loops, suitable techniques can be developed
in the same spirit. Using coordinate-space methods, analytic results are also available for higher 
orders~\cite{Luscher:1995np,Luscher:1995zz,Caracciolo:2001ki}.
Finally, mean-link improvement schemes~\cite{Lepage:1993xa} are straightforward to implement once the 
corresponding Feynman diagram is written down.
\subsection{The gauge fixing and the gluon propagator}
\label{subsec:gaugefixing}
For perturbation theory it is necessary to choose a 
gauge fixing and proceed to define the other parts of the action needed to write the Feynman rules.
In analogy with the isotropic case we choose the gauge-fixing term such that 
\be
	S_{gf} = \frac{1}{\alpha} \sum_x {\rm Tr} \left(\partial^{L}_{\mu} A_{\mu} \right)^2 \, ,
\ee
where 
$\partial^{L}_{\mu} \varphi(x) = 1/(a\, \xi_{\mu}) \left(\varphi(x)-\varphi(x - a\, \xi_{\mu}\,\hat{\mu})\right)$. 
It is straightforward to show that  
\be
	S_{gf} = \frac{1}{\tilde{\alpha}}\sum_n {\rm Tr}{\cal F}^2 \, \mbox{ for } 
		{\cal F}=b^{(l)}_\mu \hat{\partial}^{L}_{\mu} \phi_\mu(n) \, ,
	\label{GF}
\ee
where now $\hat{\partial}^{L}_{\mu} \varphi(n) = \varphi(n)- \varphi(n - \hat{\mu})$ and 
\be
	b^{(l)}_{\mu}= \left\{ \begin{array}{lr}
        		      \xi^{l+1}          & \mbox{$\mu$ fine }  \, , \\
		              \xi^{l-1}          & \hspace{3em}\mbox{$\mu$ coarse} \, , \end{array} \right.
	\label{ck}
\ee
gives the desired form once $\alpha = \xi^{-2l+d-2} \tilde{\alpha}$. The choice $l=(d-2)/2$ would be easiest
but we shall see in the following it is not the most convenient. 
The gluon propagator can now be calculated from the two field terms in Eq.~(\ref{eq1})
and the definition of the gauge fixing in Eq.~(\ref{GF}). The functional form is similar to
the isotropic case, namely
\begin{eqnarray}
\parbox{30mm}{\begin{fmfgraph*}(60,10)\fmfkeep{gl_prop}\fmfpen{thin}\fmfleft{i1}\fmfright{o1}\fmflabel{$\mu,a$}{i1}\fmflabel{$\nu,b$}{o1}\fmf{gluon,label=$k$,label.side=left}{i1,o1}\end{fmfgraph*}}
     &=&{{\delta^{ab}}\over{\hat{k}^2}}\left(\delta_{\mu\nu}-(1-\alpha)\frac{\hat{k}_{\mu} \hat{k}_{\nu}}
	{\hat{k}^2}\right) \, ,
	\label{g_pr}
\end{eqnarray}
where 
\be
	\hat{k}_{\mu} = \frac{2}{a \xi_{\mu}} \sin({\frac{a \xi_{\mu}}{2} k_{\mu}}) \, , \; \; \; 
	\hat{k}^2 = \sum_{\mu} \hat{k}^2_{\mu} \, , \; \; \; k_{\mu} \in 
	\left( - \frac{\pi}{a \xi_{\mu}}, \frac{\pi}{a \xi_{\mu}} \right) \, ,
\ee
are the dimensionful lattice momenta spanning the anisotropic Brillouin zones. All the anisotropy is 
now encapsulated in this form of the Brillouin zone. This will prove to be a constant pattern: 
for any vertex or propagator which has a continuum analogue, the form in terms of 
the (anisotropic) dimensionful momenta will resemble the isotropic ones, 
since the asymmetry now lies solely in the different momenta cut-offs.
\subsection{The Haar measure}
\label{subsec:haar}
On the lattice the functional integral is obtained by integrating over the dimensionless gauge links, 
but to do perturbation theory it must be expressed in terms of the 
dimensionful gauge fields. The Jacobian resulting from the expression of 
the Haar measure in terms of Eq.~(\ref{phield}) is~\cite{Book:Rothe}
\be
	J = \sqrt{\det\left(\frac{1}{2} M M^{\dagger}\right)} \, , \; \; \; M(\tilde{\phi}_{\mu}(n)) =
	{{1-e^{- i \tilde{\phi}_{\mu}(n)}}\over{i \tilde{\phi}_{\mu}(n)}} \, , \; \; \;
	\tilde{\phi}_{\mu}(n) = \phi^a_\mu(n) t^a \, ,
	\label{Mop}
\ee
where the $\phi^a_\mu(n)$ are the same as in Eq.~(\ref{phield}) and the $t^a$ are the adjoint generators of $SU(N_c)$. 
Reexpressing $J$ as $e^{-S_{\rm meas}}$ gives
\be
	S_{\rm meas} = -\frac{1}{2}\sum_{n,\mu}{\rm Tr} \log\left({2\left(1- \cos\tilde{\phi}_{\mu}(n)
			\right)}\over{\tilde{\phi}^2_{\mu}(n)}\right) \, .
	\label{meas}
\ee
When expanded to lowest order in the fields (which is all that is needed for 
one-loop two-point function calculations) this reads
\be
	S_{\rm meas} \simeq \frac{1}{4!} \sum_{n,\mu} N_c \delta^{ab} 
				\phi^a_{\mu}(n) \phi^b_{\mu}(n) ,
\ee
which putting back the dimensionful fields gives the vertex
\begin{eqnarray}
\parbox{30mm}{\begin{fmfgraph*}(60,10)\fmfkeep{meas_ver}\fmfpen{thin}\fmfleft{i2}\fmfright{o2}\fmflabel{$\mu,a$}{i2}\fmflabel{$\nu,b$}{o2}\fmfv{decor.shape=cross}{v2}\fmf{gluon,label=$k$,label.side=left}{i2,v2}\fmf{gluon,label=$k'$,label.side=left}{v2,o2}\end{fmfgraph*}}
    &=& -(2\pi)^4 \delta^4 (k + k')\frac{g^2}{a^2}\frac{N_c}{12}\delta_{\mu\nu}\delta^{ab}
	\frac{\xi_\mu^2}{\xi^d} \, .
	\label{ms_v} 
\end{eqnarray}
By convention all gluon momenta are incoming.
\subsection{Fadeev-Popov ghost fields}
\label{subsec:ghosts}
The Faddeev-Popov determinant, which forces the integration only on a section 
of the gauge orbits, must also be included in the action. Using ${\cal F}$ in 
Eq.~(\ref{GF}) to enforce the gauge condition on $\phi_\mu$ and using the response of the gauge fields 
on the lattice to (infinitesimal) gauge transformations~\cite{Book:Rothe} 
\be
	\hat{D}_{\mu}(\phi)= M^{-1}(\tilde{\phi}_{\mu}) \hat{\partial}^R_{\mu} + i \tilde{\phi}_{\mu} \, ,
\ee
where $\tilde{\phi}_\mu$ and the inverse adjoint-valued operator, $M$ are given 
in Eq.~(\ref{Mop}), we obtain
\be
	S_{\rm gh} = -\sum_{n,\mu} \bar{\hat{c}}^a(n) b^{(l)}_{\mu}\hat{\partial}^L_{\mu}
			\hat{D}^{ab}_{\mu}(\phi)\hat{c}^b(n) \, ,
	\label{ghost}
\ee
where $\hat{c}$ and $\bar{\hat{c}}$ are the dimensionless lattice ghost fields, 
introduced to make Eq.~(\ref{ghost}) local. From the expansion of $M^{-1}$
\be
	M^{-1}(\tilde{\phi}_{\mu}) = {\mathbb{I}}_{\rm adj} + \frac{i}{2} 
				     \tilde{\phi}_{\mu} - \frac{1}{12}\tilde{\phi}^2_{\mu} + \ldots
\ee
and by reintroducing dimensions, Eq.~(\ref{ghost}) can be written as 
\begin{eqnarray}
	S_{\rm gh} & \simeq & -\xi^{l-1} \sum_x\left( \bar{c}^a (x)\delta^{ab}\partial^L_{\mu}
				\partial^R_{\mu}c^b(x) + gf^{abc}\bar{c}^a (x)\partial^L_{\mu}
				\lbrack A^c_{\mu}(x)(1 + \frac{a\, \xi_{\mu}}{2}\partial^R_{\mu})
				\rbrack c^b(x)\right. \nonumber \\
		    & &       + \left.\frac{1}{2!}\frac{g^2 a^2 \xi_{\mu}^2}{12}\delta_{\mu \nu}
				\lbrace t^c, t^d \rbrace_{ab} (\partial^R_{\mu}\bar{c}^a (x))
				(\partial^R_{\mu}c^b(x)) A^c_{\mu}(x)A^d_{\nu}(x) \right) \, , \nonumber
\end{eqnarray}
by including only terms which are relevant for one-loop two-point functions.
The factor $\xi^{l-1}$ at the front of $S_{\rm gh}$ gives rise to spurious 
coefficients with no continuum analogue which however, cancel in 
any graph with no ghost outer legs. Alternatively the coefficients can be absorbed in a  
redefinition of the ghost fields, $\xi^{(l-1)/2} c \to c$ or 
by setting $l=1$, thus rescaling the lattice 
gauge-fixing parameter $\tilde{\alpha} = \xi^{4-d} \alpha$. For each of these alternatives the ghost propagator and the two ghost one- and two-gluon 
vertices are 
\begin{eqnarray}
\parbox{35mm}{\begin{fmfgraph*}(60,10)\fmfkeep{gh_prop}\fmfpen{thin}\fmfleft{igh}\fmfright{ogh}\fmflabel{$a$}{igh}\fmflabel{$b$}{ogh}\fmf{ghost,label=$k$,label.side=left}{igh,ogh}\end{fmfgraph*}} 
     &=&\delta^{ab} \, \frac{1}{\hat{k}^2} \label{gh}  , \\
     & & \nonumber \\
\parbox{35mm}{\begin{fmfgraph*}(55,40)\fmfkeep{gh_2gl_ver}\fmfpen{thin}\fmfleft{i31,i32}\fmfright{o3}\fmflabel{$p,a$}{i31}\fmflabel{$p',b$}{i32}\fmflabel{$\mu,c$}{o3}\fmf{gluon,label=$k$,label.side=left}{v3,o3}\fmf{ghost}{i31,v3}\fmf{ghost}{v3,i32}\end{fmfgraph*}}
     &=& i g (2 \pi)^4 \delta^4(k + p - p') f^{abc}{\hat{p}'}_{\mu}\, \tilde{p}_{\mu} \label{g_gh} , \\
     & & \nonumber \\
     & & \nonumber \\
\parbox{35mm}{\begin{fmfgraph*}(55,40)\fmfkeep{2gh_2gl_ver}\fmfpen{thin}\fmfleft{i41,i42}\fmfright{o41,o42}\fmflabel{$p,b$}{i42}\fmflabel{$k,\mu,c$}{i41}\fmflabel{$p',a$}{o42}\fmflabel{$k',\nu,d$}{o41}\fmf{gluon}{i41,v4}\fmf{gluon}{v4,o41}\fmf{ghost}{i42,v4}\fmf{ghost}{v4,o42}\end{fmfgraph*}}
     &=&\frac{1}{12}g^2 a^2 (2\pi)^4 \delta^4(k+k'+p-p')\lbrace t^c, t^d \rbrace_{ab} \delta_{\mu\nu}\,
	\xi_{\mu}^2 \,\hat{p}_{\mu}\, {\hat{p}'}_{\nu} ,
	\label{2g_2gh} \\
     & & \nonumber 
\end{eqnarray}
with $\tilde{p}_{\mu} = \cos{\frac{1}{2}p_{\mu}a\xi_{\mu}}$. As usual
the vertex which has a continuum analogue carries no explicit dependence on the anisotropy.
\subsection{Gluon vertices}
\label{subsec:gluonvertices}
Taking into account the three-gluon terms arising from Eq.~(\ref{eqn:S-Wilson}), 
one of which comes from the irrelevant operator ${\cal O}$ defined in Eq.~(\ref{gluon}) gives
\be
	S^{(3)}_{\rm W} = g \sum_x f^{abc} \delta_{\lambda\nu}\left(A^a_{\mu}(x)+\frac{a \,\xi_{\mu}}{2} 
				\partial^R_{\nu} A^a_{\mu}(x)\right)(\partial^R_{\mu}A^b_{\nu}(x))
				A^c_{\lambda}(x) ,
\ee
which, after a Fourier transformation, leads to an expression equivalent to the isotropic case
\vspace{4mm}
\begin{eqnarray}
\parbox{35mm}{\begin{fmfgraph*}(55,40)\fmfkeep{3gh_ver}\fmfpen{thin}\fmfleft{i51,i52}\fmfright{o5}\fmflabel{$k,\mu,a$}{i52}\fmflabel{$k',\nu,b$}{i51}\fmflabel{$k'',\lambda,c$}{o5}\fmf{gluon}{i51,v5}\fmf{gluon}{o5,v5}\fmf{gluon}{i52,v5}\end{fmfgraph*}}
    &=& i g (2\pi)^4 \delta^4 (k+k'+k'') f^{abc}\left(\delta_{\nu\lambda}\widehat{(k''-k')}_{\mu}
	\tilde{k}_{\nu}\right.\nonumber\\
    & & +\left.\delta_{\mu\lambda} \widehat{(k-k'')}_{\nu} \tilde{k'}_{\lambda}+\delta_{\mu\nu}
	\widehat{(k'-k)}_{\lambda} \tilde{k''}_{\mu}\right) .
	\label{3g_v}
\end{eqnarray}
Taking the four-gluon terms arising from Eq.~(\ref{eqn:S-Wilson}), mostly coming from ${\cal O}$, 
(see Ref.~\onlinecite{Book:Rothe} for details), 
and rechecking the cancellations which must occur due to Bose symmetry, one 
finds the four-gluon vertex whose expression is given in 
Appendix~\ref{app:4glver}.
\subsection{One-loop correction vertex}
\label{subsec:1loopcorrection}
The one-loop corrections to the action coming from Eqs.~(\ref{cs}) and~(\ref{cs1}) give rise to an extra 
vertex which reads
\vspace{4mm}
\begin{eqnarray}
\parbox{30mm}{\begin{fmfgraph*}(60,10)\fmfkeep{corr_ver}\fmfpen{thin}\fmfleft{ic}\fmfright{oc}\fmflabel{$\mu,a$}{ic}\fmflabel{$\nu,b$}{oc}\fmfv{decor.shape=circle}{vc}\fmf{gluon,label=$k$,label.side=left}{ic,vc}\fmf{gluon,label=$k'$,label.side=left}{vc,oc}\end{fmfgraph*}}
  &=&  - (2 \pi)^4 \delta^4 (k+k') g^2  \delta^{ab}  (\delta_{\mu\nu} \sum_{\rho} \eta^{(1)}_{\mu\rho}
	\hat{k}_{\rho}^2-\eta^{(1)}_{\mu\nu}\hat{k}_{\mu}\hat{k}_{\nu}) ,
	\label{v_c}
\end{eqnarray}
where 
\be
	  \eta^{(1)}_{\mu\nu}= \left\{ \begin{array}{lr}
	              \eta_{f\!f}^{(1)} & \mbox{$\mu,\nu$ fine }\\
        	      \eta_{cf}^{(1)}  &\hspace{3em} \mbox{$\mu$ coarse(fine) and $\nu$ fine(coarse)}\\
	              \eta_{cc}^{(1)} & \mbox{$\mu,\nu$ coarse}
	\end{array} \right.\label{v_c1}
\ee
In the 3+1 case one can either set $\eta_{f\!f}^{(1)}$ to zero or leave it free,
as by the traceless property of the vertex it will always cancel.
\subsection{The continuum limit and anisotropic renormalization}
\label{subsec:ContinuumLimit}
The calculation of  the one-loop corrections to the gluon self-energy, 
$\Sigma_{\mu \nu,ab}(p)$, using the Wilson action involves five Feynman diagrams~\cite{Kawai:1981ja} and 
Eq.~(\ref{v_c}).
Each diagram is a function of the external momenta $p$ and can be written as
\be
	G(p) = \int {d^4 k\over (2 \pi)^4} F(k,p) \, ,
	\label{eq2.1}
\ee
where $k$ is the integration momenta.
Since we are interested in the continuum limit of Eq.~(\ref{eq2.1}), if the integral is 
ultraviolet convergent we can simply substitute the function $F(k,p)$ with its continuum equivalent.
Otherwise, if Eq.~(\ref{eq2.1}) is divergent and contains only massive propagators so that 
$F(k,p)$ is finite for any set of momenta going to zero, one can use the lattice 
version of the BPHZ technique~\cite{Book:Collins,Reisz:1988pw,Reisz:1988da} by writing 
\begin{eqnarray}
	G(p) &=& \int {d^4 k\over (2 \pi)^4}  
		 \bigg[ F(k,p) - \sum_{n=0}^{n_F} 
			     {1\over n!} \, p_{\mu_1} \ldots p_{\mu_n} 
			     \bigg( {\partial\over \partial p_{\mu_1} } \ldots
		            {\partial\over \partial p_{\mu_n}} F(k,p) \bigg)_{p=0} \bigg] \label{eq2.2}\\ 
	                &&   + \int {d^4 k\over (2 \pi)^4} \sum_{n=0}^{n_F} 
			     {1\over n!} \, p_{\mu_1} \ldots p_{\mu_n} 
			     \bigg( {\partial\over \partial p_{\mu_1} } \ldots
		            {\partial\over \partial p_{\mu_n}} F(k,p)\bigg)_{p=0} \nonumber \\ 
	  &\equiv& G^c(p) +\, G^L (p) \, , \nonumber 
	\end{eqnarray}
where $n_F$ is the degree of divergence of the diagram.
The first term in Eq.~(\ref{eq2.2}) is ultraviolet 
finite~\cite{Reisz:1988pw,Reisz:1988da} and therefore its continuum limit can be taken, 
making it independent of $\xi$. All the effects of the lattice regularization
remain in the second term, which is simply a polynomial in the 
external momenta with coefficients given by zero-momentum lattice integrals. 

On the other hand, if a diagram contains massless propagators, as in our case, 
more care is needed: indeed an expansion around $p=0$ can give rise to infrared divergences. 
A simple recipe is to introduce an intermediate infrared regularization. Given an anisotropic
cut-off the introduction of a mass, $m$ in the propagators is the most suitable 
solution~\cite{Luscher:1995np,Caracciolo:1992cp,Burgio:1996ji}, while
dimensional regularization is popular in the
literature for the isotropic case~\cite{Kawai:1981ja,Alles:1997cy}.
$G^c(p)$ and $G^L(p)$ are then divergent for 
$m\to 0$ separately but the divergences cancel in the sum.

\section{Results}
\label{sec:results}
In this section we present the results of the one-loop correction to the gluon propagator using the Wilson 
action in the Feynman gauge for a general anisotropic lattice in four dimensions. 
Applying the procedure explained in 
Section~(\ref{subsec:lorentzinvariance}) we calculate the values of the coefficients 
which restore on-shell Lorentz 
invariance, where the definition and treatment of the lattice integrals $\B$ is given in 
Appendix~(\ref{app:Integrals}). 
\subsection{On-shell Lorentz invariance}
\label{subsec:lorentzinvariance}
As explained in Section~\ref{sec:aniso}, the one-loop propagator obtained 
from the Feynman rules given in Section~\ref{sec:pert} will not in general 
satisfy Lorentz invariance. 
The free parameters in Eq.~(\ref{v_c1}) must be tuned to restore 
the symmetry~\cite{Karsch:1982ve,GarciaPerez:1997ft,Drummond:2002yg}. 
From Eq.~(\ref{v_c}) it is clear that not every Lorentz-breaking 
term can be cancelled, as expressions of the form $\delta_{\mu\nu}/a^2$ and 
$\delta_{\mu\nu} p_{\mu}^2$ arise from each diagram. The first non-trivial 
result of our calculation is that, for any value of $d$ and $\xi$,
such terms cancel exactly when the diagrams are summed up, just as in 
the isotropic case, independently of the tuning procedure.

The remaining Lorentz-breaking artefacts arising from 
the mixing of longitudinal and transverse field components have the correct
structure and can be cured by tuning. 
We choose to fix the parameters in the action by demanding the recovery of Lorentz invariance for 
on-shell physical soft gluon modes, imposing
that the two physical eigenvalues of the one-loop propagator 
vanish for $E^2=p^2$.
Since this is a gauge-invariant condition we restrict ourselves to a 
particular gauge. We have chosen the Feynman gauge for which $\alpha=1$. 
By injecting the gluon momentum in any possible direction and calculating the
eigenvalues and eigenvectors of the propagator we obtain a general
condition for $\eta_{cc}^{(1)} - \eta_{cf}^{(1)}$ and $\eta_{f\!f}^{(1)} - \eta_{cf}^{(1)}$, 
independent of the direction of $p_{\mu}$. 
The value of $\eta_{cf}^{(1)}$ remains unconstrained as it can be reabsorbed
in a one-loop $\beta$-shift. 

To identify the physical eigenmodes, the momenta directions with residual symmetry are first identified;
In the case of the 2+2 lattice, this corresponds to any momenta in the coarse-coarse or fine-fine planes. Then
for more general momenta, the axis of propagation was smoothly varied away from these symmetric cases, and the 
eigenmodes continuously traced. This investigation lead to a generalisation of the polarisation condition, 
$p_\mu \epsilon_\mu=0$ for on-shell gluon polarisation vector $\epsilon_\mu$. The resulting 2+2 lattice 
polarisation condition is 
\be
     p_\mu Z_{\mu\nu}\epsilon_\nu=0,
\ee
with $Z_{\mu\nu}$ is the diagonal matrix $Z_{\mu\nu} = \delta_{\mu\nu} \left(Z_c \delta_{\mu c} + 
Z_f \delta_{\mu f} \right)$ and $Z_c$, $Z_f$ the coarse and fine gluon field renormalisation coefficients.
$\delta_{\mu f}$ and $\delta_{\mu c}$ are one(zero) if $\mu$ is fine(coarse).

\subsection{The one-loop coefficients for the restoration of Lorentz invariance}
\label{subsec:oneloopcoeffs}
For 2+2 anisotropic lattice the  one-loop coefficients for the restoration of Lorentz invariance are
\begin{eqnarray}
	\eta_{cc}^{(1)} - \eta_{cf}^{(1)} &=&
			-\frac{1}{2 N_c}\Bigg[ {\cal B}_\xi^c(1,1)-\frac{1}{4}\Bigg]+N_c\Bigg[ -\frac{1}{16} + 
			\frac{{\cal B}_\xi(1)}{6}\bigg( \frac{7}{2} + \frac{1}{\xi^2} \bigg) +
			\frac{{\cal B}_\xi^c(1,1)}{4} \nonumber \\ 
					  & & -\frac{{\cal B}_\xi^c(2,1)}{3} \bigg(2 + \frac{5}{2\xi^4} +
			  \frac{11}{2 \xi^2} \bigg) + \frac{{\cal B}_\xi^f(2,1,1)}{6} \bigg(\frac{1}{2} + 
              		 \xi^2 \bigg) \Bigg] \, , 
	\label{coeff22a} 
\end{eqnarray}
and
\begin{eqnarray}
	\eta_{f\!f}^{(1)} - \eta_{cf}^{(1)} &=&
		\frac{1}{2 N_c} \bigg[ \frac{1}{4}-\frac{1}{2 \xi^2} +\frac{{\cal B}_\xi^c(1,1)}{\xi^4}\bigg] - 
			\frac{N_c}{2} \bigg[ \frac{1}{4} \bigg( \frac{1}{2} - \frac{1}{\xi^2}\bigg) + 
			{\cal B}_\xi(1) \bigg(\frac{1}{2} + \frac{1}{3 \xi^2} \bigg) \nonumber \\ 
			 	            & & + \frac{{\cal B}_\xi^c(1,1)}{2 \xi^4} - 
			{\cal B}_\xi^c(2,1) \bigg( \frac{5}{3 \xi^4} + \frac{1}{\xi^2} + \frac{1}{6} \bigg)-
			\frac{{\cal B}_\xi^f(2,1,1)}{3 \xi^2} \bigg] \, ,
		\label{coeff22b}
\end{eqnarray}
which diverge logarithmically with $\xi$, while for 3+1 we have 
\begin{eqnarray}
	\eta_{cc}^{(1)} - \eta_{cf}^{(1)} &=&
		\frac{N_c}{\xi} \Bigg[ \frac{{\cal B}_\xi(1)}{6} \bigg(\frac{\xi^2}{3} + \frac{19}{6} + 
		\frac{7}{2 \xi^2}\bigg) + \frac{{\cal B}_\xi^c(1,1)}{4} \bigg(1 + \frac{1}{\xi^2} \bigg)
		\nonumber \\ 
		& & -{\cal B}_\xi^c(2,1) \bigg(\frac{1}{3}+\frac{11}{6 \xi^2} +\frac{5}{2 \xi^4} \bigg) 
		    -\frac{1}{8}\Bigg] - \frac{1}{2 \xi N_c} \Bigg[{\cal B}_\xi^c(1,1)\bigg(1+\frac{1}{\xi^2} 
		    \bigg) - \frac{1}{2} \Bigg] \,,
	\label{coeff31}
\end{eqnarray}
which agrees with the general result of Ref.~\onlinecite{GarciaPerez:1997ft} 
and the $N_c=3$ result of Ref.~\onlinecite{Drummond:2002yg}. 
The techniques used by both groups
involve extrapolations in some suitable parameter, absent in our treatment. 
The functions $\B$ are defined in Appendix~\ref{app:Integrals}. Using the
results given there the Hamiltonian limit, 
$\xi \to \infty$ can be treated analytically also and agrees again with 
Ref.~\onlinecite{GarciaPerez:1997ft}. The difference in the parameters, 
$\eta_{cc}^{(1)}-\eta_{cf}^{(1)}$ and $\eta_{f\!f}^{(1)} - \eta_{cf}^{(1)}$ appears as a polynomial
in $N_c$ with terms proportional to $1/N_c$ and $N_c$ only, as expected. 
Numerical values for the coefficients in
this polynomial for a range of anisotropies are given to ten decimal places in 
Tables~(\ref{chi-coeff22},\ref{chi-coeff31}).

\subsection{The gluon self-energy in the Feynman gauge}
\label{subsec:gluonselfenergy}
Once the values in Eqs.~(\ref{coeff22a}),~(\ref{coeff22b}) and~(\ref{coeff31}) 
are inserted, the one-loop correction to the gluon self-energy reads
\be
	\Sigma^{(1)}_{\mu \nu,ab}(p) = g^2 \delta^{ab} \left( \delta_{\mu \nu} p^2 - 
					p_\mu p_\nu \right)\bigg\{ A - \eta_{cf}^{(1)} - B\bigg(
			\delta_{\mu c} \delta_{\nu c} -\delta_{\mu f} \delta_{\nu f} \bigg) \bigg\} \, .
	\label{propcom}
\ee
For 2+2 we have 
\begin{eqnarray}
	A &=& -\frac{1}{8N_c} + N_c\Bigg[\frac{1}{16} + \frac{{\cal B}_\xi(1)}{2} \bigg(\frac{1}{3} 
		- \frac{1}{4 \xi^2} \bigg) + \frac{{\cal B}_\xi^c(2,1)}{4 \xi^2} 
		\bigg(\frac{7}{3 \xi^2} - 1 \bigg) 
		-\frac{{\cal B}_\xi^f(2,1,1)}{12} \nonumber \\
	  &  & + \frac{5}{3} X(\xi) \Bigg] \\
	B &= & N_c \Bigg[ \frac{{\cal B}_\xi(1)}{12} \bigg( 1 + \frac{1}{2 \xi^2} \bigg) 
		- \frac{{\cal B}_\xi^c(2, 1)}{4 \xi^2} \bigg( 1 + \frac{1}{\xi^2} \bigg) \Bigg] ,
	\end{eqnarray}
while for 3+1 
\begin{eqnarray}
	A &=& \frac{1}{2 N_c \xi} \Bigg[ \frac{{\cal B}_\xi^c(1,1)}{\xi^2} - \frac{1}{2} \Bigg] + 
	      \frac{N_c}{\xi} \Bigg[ \frac{1}{8} + \frac{5 \xi}{3} X(\xi) - 
		\frac{{\cal B}_\xi^c(1,1)}{4 \xi^2} + \frac{{\cal B}_\xi^c(2,1)}{4 \xi^2} 
		\bigg( \frac{7}{\xi^2} + \frac{1}{3} \bigg) \nonumber \\ 
	  & & + \frac{{\cal B}_\xi(1)}{6} \bigg( \frac{1}{3} - \frac{5}{2 \xi^2} \bigg) \Bigg] \, ,\\
	B &=& \frac{N_c}{\xi} \Bigg[ \frac{{\cal B}_\xi(1)}{6} \bigg( \frac{1}{2} + 
		\frac{1}{\xi^2} \bigg) - \frac{{\cal B}_\xi^c(2,1)}{4\xi^2}\bigg(1+\frac{3}{\xi^2}\bigg)
		\Bigg] \, ,
\end{eqnarray}
where
\be
	X(\xi) = \frac{1}{16 \pi^2} (-\log(a^2 p^2) + F_0(\xi) - \gamma_E + \frac{28}{9}) \, .  
\ee
We stress that the term proportional to $B$ in 
Eq.~(\ref{propcom}) does not break Lorentz invariance but arises from the difference in the 
renormalizations of the fine and coarse fields $A_{\mu}$. This difference must be taken into account
when calculating physical quantities such as the $\Lambda$ parameter or matrix elements. After  
including the difference, the physical modes will still be transverse. 

Setting $\xi =1$ we recover from Eq.~(\ref{propcom}) the well-known value of 
the gluon self-energy~\cite{Kawai:1981ja,Luscher:1995np,Alles:1997cy}
\begin{eqnarray}
	\Sigma_{\mu \nu,ab}(p) &=& \delta^{ab}(\delta_{\mu \nu}p^2-p_\mu p_\nu)\bigg[ 
                		   1-g^2 \bigg\{\frac{N_c}{16 \pi^2}\bigg( -\frac{5}{3}
				   \log(p^2 a^2) + \frac{28}{9} \bigg) - \frac{1}{8 N_c} \nonumber \\ 
		               & & + N_c \bigg( \frac{7}{72} Z_0 + \frac{5}{48 \pi^2}(F_0 - \gamma_E) + 
			          \frac{1}{16} \bigg) \bigg\} \bigg] + O(g^4) \, ,
	\label{prop}
\end{eqnarray}
where $Z_0$ and $F_0$ are standard integrals in perturbation theory on the isotropic lattice, and are defined in Ref.~\onlinecite{Caracciolo:1992cp}. 
In order to calculate coefficients at the one-loop level in mean-link improved perturbation theory, 
the evaluation of either the plaquette, $\langle{\rm Tr }P_{\mu\nu}\rangle$ or the link trace in Landau gauge, 
$\langle{\rm Tr}\,U_\mu\rangle$ is required. We find
\begin{eqnarray}
	\frac{1}{N_c}\langle{\rm Tr}\,P_{cc}\rangle &=& 1-\frac{g^2}{2 \xi^{d - 2}}(N_c-\frac{1}{N_c})\B^c(1, 1) \, ,\\
	\frac{1}{N_c}\langle{\rm Tr}\,P_{fc}\rangle &=& 1-\frac{g^2}{4\xi^{d}}(N_c-\frac{1}{N_c})
					  \left(\xi^2 \B^f(1,1)+\B^c(1, 1)\right) \, ,\\
	\frac{1}{N_c}\langle{\rm Tr}\,P_{f\!f}\rangle &=& 1-\frac{g^2}{2 \xi^{d}}(N_c-\frac{1}{N_c})\B^f(1, 1) \, ,
\end{eqnarray}
and
\begin{eqnarray}
	\frac{1}{N_c}\langle{\rm Tr}\,U_c\rangle &=& 1-\frac{g^2}{4 \xi^d}(N_c-\frac{1}{N_c})\left[\xi^2 \B(1) -
					\B^c(2, 1)\left(1 -\frac{1}{\alpha}\right)\right] \, ,\\
	\frac{1}{N_c}\langle{\rm Tr}\,U_f\rangle &=& 1-\frac{g^2}{4 \xi^d}(N_c-\frac{1}{N_c})\left[\B(1) - 
				       \B^f(2, 1)\left(1 -\frac{1}{\alpha}\right)\right] \, ,
\end{eqnarray}
where $\B^f(q,1)$ satisfies Eq.~(\ref{rel1}). Numerical values of the
$N_c$-polynomial coefficients, $\eta^{(1)}_{cc}$, $\eta^{(1)}_{cf}$ and $\eta^{(1)}_{f\!f}$ 
for a range of anisotropies and including these improvement terms are given in 
Tables~(\ref{chi-coeff22-mp},\ref{chi-coeff22-ml},\ref{chitadpole=coeff31},\ref{chimean=coeff31}).
\begin{table}[h]
\begin{tabular}{cccccc}
\hline
$\xi$ & \multicolumn{2}{c}{$\eta_{cc}^{(1)}-\eta_{cf}^{(1)}$}
      &\rule{0.5em}{0em}& \multicolumn{2}{c}{$\eta_{f\!f}^{(1)}-\eta_{cf}^{(1)}$} \\
\hline
      &     $1/N_c$   &      $N_c$      &&     $1/N_c$   &      $N_c$ \\
\hline
 1 &           0      &        0        &&         0       &        0 \\
 2 & $ -0.1161042701 $&$ 0.0958601237 $ && $ 0.0775690169 $&$ -0.0741548219  $ \\
 3 & $ -0.1886258210 $&$ 0.1529850132 $ && $ 0.1010941459 $&$ -0.1044390683 $ \\
 4 & $ -0.2391230676 $&$ 0.1920838712 $ && $ 0.1107973557 $&$ -0.1217185361 $ \\
 5 & $ -0.2774799637 $&$ 0.2215115752 $ && $ 0.1156439679 $&$ -0.1335092263 $ \\
 6 & $ -0.3083219799 $&$ 0.2450412380 $ && $ 0.1183899089 $&$ -0.1424129673 $ \\
 7 & $ -0.3340937184 $&$ 0.2646297423 $ && $ 0.1200891686 $&$ -0.1495669092 $ \\
 8 & $ -0.3562244833 $&$ 0.2814071688 $ && $ 0.1212112364 $&$ -0.1555537098 $ \\
 9 & $ -0.3756169823 $&$ 0.2960809268 $ && $ 0.1219898822 $&$ -0.1607076435 $ \\
10 & $ -0.3928761275 $&$ 0.3091218866 $ && $ 0.1225517876 $&$ -0.1652371442 $ \\
\hline
\end{tabular}
\caption{$N_c$-polynomial coefficients of $\eta_{cc}^{(1)}-\eta_{cf}^{(1)}$ and $\eta_{f\!f}^{(1)}-\eta_{cf}^{(1)}$ 
for $\xi=1,\cdots,10$ on a 2+2 anisotropic lattice. }
\label{chi-coeff22}
\end{table}
\begin{table}
\begin{tabular}{ccc}
\hline
$\xi$ & \multicolumn{2}{c}{$\eta_{cc}^{(1)}-\eta_{cf}^{(1)}$} \\
\hline
      &   $1/N_c$   & $N_c$  \\ 
\hline
 1 & 0  & 0 \\
 2 & -0.0853420430 & 0.0663515978 \\
 3 & -0.1202379609 & 0.0892976526 \\
 4 & -0.1389282597 & 0.1007025039 \\
 5 & -0.1504977336 & 0.1074686585 \\
 6 & -0.1583451793 & 0.1119335310 \\
 7 & -0.1640119208 & 0.1150960459 \\
 8 & -0.1682939409 & 0.1174518583 \\
 9 & -0.1716426764 & 0.1192740369 \\
10 & -0.1743328788 & 0.1207251404 \\
\hline
\end{tabular}
\caption{$N_c$-polynomial coefficients of $\eta_{cc}^{(1)}-\eta_{cf}^{(1)}$ for $\xi=1,\cdots,10$ 
on a 3+1 anisotropic lattice. }
\label{chi-coeff31}
\end{table}
\begin{table}
\begin{tabular}{cccccc}
\hline
$\xi$ & 
\multicolumn{2}{c}{$\eta_{cc}^{(1)}-\eta_{cf}^{(1)}$}
&\rule{0.5em}{0em} & \multicolumn{2}{c}{$\eta_{f\!f}^{(1)}- \eta_{cf}^{(1)}$} \\
\hline
 & $     1/N_c$    &     $N_c$    &&     $1/N_c$  & $N_c$ \\
\hline
 1 & $ -1/16 $     & $  1/16 $    && $  1/16 $    & $ -1/16 $   \\
 2 & -0.2366564052 & 0.2164122588 && 0.1981211519 & -0.1947069547 \\
 3 & -0.3454387314 & 0.3097979237 && 0.2579070564 & -0.2612519788 \\
 4 & -0.4211846015 & 0.3741454050 && 0.2928588896 & -0.3037800700 \\
 5 & -0.4787199456 & 0.4227515571 && 0.3168839498 & -0.3347492082 \\
 6 & -0.5249829699 & 0.4617022280 && 0.3350508989 & -0.3590739573 \\
 7 & -0.5636405777 & 0.4941766016 && 0.3496360278 & -0.3791137684 \\
 8 & -0.5968367250 & 0.5220194105 && 0.3618234781 & -0.3961659514 \\
 9 & -0.6259254734 & 0.5463894180 && 0.3722983733 & -0.4110161346 \\
10 & -0.6518141912 & 0.5680599503 && 0.3814898513 & -0.4241752080 \\
\hline
\end{tabular}
\caption{$N_c$-polynomial coefficients of $\eta_{cc}^{(1)}-\eta_{cf}^{(1)}$ and $\eta_{f\!f}^{(1)}-\eta_{cf}^{(1)}$ for $\xi=1,\cdots,10$ on a 
2+2 anisotropic lattice using mean (coarse) plaquette improvement.}
\label{chi-coeff22-mp}
\end{table}
\begin{table}
\begin{tabular}{cccccc}
\hline
$\xi$ & \multicolumn{2}{c}{$\eta_{cc}^{(1)}-\eta_{cf}^{(1)}$}
&\rule{0.5em}{0em} & \multicolumn{2}{c}{$\eta_{f\!f}^{(1)}-\eta_{cf}^{(1)}$} \\
\hline
 & $     1/N_c$    &     $N_c$    &&     $1/N_c$  & $N_c$ \\
\hline
1  & $ -3Z_0/8$ & $ 3Z_0/8$ && $ 3Z_0/8$ & $ -3Z_0/8$  \\
2  & -0.2304051667 & 0.2101610203 && 0.1918699134 & -0.1884557162 \\
3  & -0.3391678963 & 0.3035270885 && 0.2516362213 & -0.2549811436 \\
4  & -0.4150758989 & 0.3680367024 && 0.2867501870 & -0.2976713674 \\
5  & -0.4727526305 & 0.4167842420 && 0.3109166347 & -0.3287818930 \\
6  & -0.5191216331 & 0.4558408912 && 0.3291895621 & -0.3532126205 \\
7  & -0.5578576084 & 0.4883936322 && 0.3438530585 & -0.3733307991 \\
8  & -0.5911125215 & 0.5162952070 && 0.3560992746 & -0.3904417479 \\
9  & -0.6202462174 & 0.5407101620 && 0.3666191173 & -0.4053368786 \\
10 & -0.6461700016 & 0.5624157607 && 0.3758456617 & -0.4185310183 \\
\hline
\end{tabular}
\caption{$N_c$-polynomial coefficients of $\eta_{cc}^{(1)}-\eta_{cf}^{(1)}$ and $\eta_{f\!f}^{(1)}-\eta_{cf}^{(1)}$ for $\xi=1,\cdots,10$ on a 2+2 anisotropic lattice using mean (coarse) link improvement. $Z_0$ is defined 
in Section~\ref{subsec:oneloopcoeffs}.}
\label{chi-coeff22-ml}
\end{table}
\begin{table}
\begin{tabular}{ccc}
\hline
$\xi$ & \multicolumn{2}{c}{$\eta_{cc}^{(1)}  - \eta_{cf}^{(1)}$}  \\
\hline
   & $ 1/N_c$ & $ N_c$     \\
\hline
 1 & $ -1/16     $ & $ 1/16$     \\
 2 & -0.1694788603 & 0.1504884150 \\
 3 & -0.2118450434 & 0.1809047350 \\
 4 & -0.2337180290 & 0.1954922732 \\
 5 & -0.2468908748 & 0.2038617997 \\
 6 & -0.2556482395 & 0.2092365912 \\
 7 & -0.2618777619 & 0.2129618870 \\
 8 & -0.2665309579 & 0.2156888753 \\
 9 & -0.2701369251 & 0.2177682856 \\
10 & -0.2730125217 & 0.2194047833 \\
\hline
\end{tabular}
\caption{$N_c$-polynomial coefficients of $\eta_{cc}^{(1)}-\eta_{cf}^{(1)}$ for $\xi=1,\cdots,10$ on a 
3+1 anisotropic lattice using mean (coarse) plaquette improvement.}
\label{chitadpole=coeff31}
\end{table}
\begin{table}
\begin{tabular}{ccc}
\hline
$\xi$ & \multicolumn{2}{c}{$\eta_{cc}^{(1)}-\eta_{cf}^{(1)}$}\\
\hline
      &  $1/N_c$         & $N_c$ \\
\hline
 1 & $ -3Z_0/8$          & $ 3Z_0/8$  \\
 2 & -0.1643550433 & 0.1453645980 \\
 3 & -0.2068380749 & 0.1758977666 \\
 4 & -0.2288323401 & 0.1906065843 \\
 5 & -0.2420838939 & 0.1990548187 \\
 6 & -0.2508913882 & 0.2044797400 \\
 7 & -0.2571539394 & 0.2082380645 \\
 8 & -0.2618297768 & 0.2109876942 \\
 9 & -0.2654518384 & 0.2130831989 \\
10 & -0.2683392407 & 0.2147315023 \\
\hline
\end{tabular}
\caption{$N_c$-polynomial coefficients of $\eta_{cc}^{(1)}-\eta_{cf}^{(1)}$ for $\xi=1,\cdots,10$ on a 
3+1 anisotropic lattice using mean (coarse) link improvement. $Z_0$ is defined in 
Section~\ref{subsec:oneloopcoeffs}.}
\label{chimean=coeff31}
\end{table}
\clearpage

\section{Conclusions}
\label{sec:conclude}
In this paper, a generalization of the Wilson discretization to an anisotropic
lattice with two coarse and two fine directions has been described.
In particular, an important distinction between the 2+2 anisotropy and the well-established 
3+1 case has been emphasized. The difference is that the
coefficients in the 2+2 action must be determined before simulation to ensure
Lorentz invariance, while any 3+1 action leads to a Lorentz-invariant theory
once the ratio of scales, $\xi$ is determined. We are currently investigating
non-perturbative techniques for computing these coefficients in the 2+2 case~\cite{Burgio:2002aa}. 

The main result of the paper was to compute these coefficients to first order in
perturbation theory. While the focus of the calculation was on determining the
Feynman rules for the 2+2 lattice Wilson gauge action, a more general
prescription was developed to allow the 3+1 case to be investigated as well. 
This allowed us to check our results against previously published work. The
results in a mean-link improvement scheme were presented. 

The usefulness of this scheme arises from the need to make accurate
calculations of form-factors and matrix elements at high momentum.  
This paper establishes the tools for perturbation theory calculations on the 
2+2 anisotropic lattice, which will be important later when computing the 
matching factors to link calculations of weak-decay matrix elements to their 
continuum counterparts. As part of this program, quark fields on 2+2 anisotropic 
lattices are under consideration.

\appendix
\section{The Four Gluon Vertex}
\label{app:4glver}
\vspace{4mm}
\begin{eqnarray}
\parbox{35mm}{\begin{fmfgraph*}(55,40)\fmfkeep{4gl_ver}\fmfpen{thin}\fmfleft{i61,i62}\fmfright{o61,o62}\fmflabel{$k,\mu,a$}{i61}\fmflabel{$q,\nu,b$}{i62}\fmflabel{$s,\rho,d$}{o61}\fmflabel{$r,\lambda,c$}{o62}\fmf{gluon}{i61,v6}\fmf{gluon}{v6,o61}\fmf{gluon}{i62,v6}\fmf{gluon}{v6,o62}\end{fmfgraph*}}
 &=&  -g^2 \left\{\sum_e f^{abe}f^{cde}\right. \left\{\delta_{\mu\lambda}
	\delta_{\nu\rho} \left[\widetilde{(q-s)}_{\mu}\widetilde{(k-r)}_{\nu}-\frac{a^4}{12}
	\xi^2_{\mu}\xi^2_{\nu}\hat{k}_{\nu}
	\hat{q}_{\mu}\hat{r}_{\nu}\hat{s}_{\mu}\right]\right. \nonumber \\
 && -\delta_{\mu\rho}\delta_{\nu\lambda}\left[\widetilde{(q-r)}_{\mu}\widetilde{(k-s)}_{\nu}-
	\frac{a^4}{12} \xi^2_{\mu}\xi^2_{\nu}\hat{k}_{\nu}\hat{q}_{\mu}\hat{r}_{\mu}\hat{s}_{\nu}\right]\nonumber \\
 && +\frac{a^2}{6}\delta_{\nu\lambda}\delta_{\nu\rho}\xi^2_{\nu}\widehat{(s-r)}_{\mu}\hat{k}_{\nu}
	\tilde{q}_{\mu}-\frac{a^2}{6}\delta_{\mu\lambda}\delta_{\mu\rho}\xi^2_{\mu}
	\widehat{(s-r)}_{\nu}\hat{q}_{\mu}\tilde{k}_{\nu} \nonumber \\
 && +\frac{a^2}{6}\delta_{\mu\nu}\delta_{\mu\rho}\xi^2_{\rho}\widehat{(q-k)}_{\lambda}\hat{r}_{\rho}
	\tilde{s}_{\lambda}-\frac{a^2}{6}\delta_{\mu\nu}\delta_{\mu\lambda}
	\xi^2_{\lambda}\widehat{(q-k)}_{\rho}\hat{s}_{\lambda}\tilde{r}_{\rho} \nonumber \\
 && +\left.\frac{a^2}{12}\delta_{\mu\nu}\delta_{\mu\lambda}\delta_{\mu\rho}\xi^2_{\mu}\sum_{\sigma}
	\widehat{(q-k)}_{\sigma}\widehat{(s-r)}_{\sigma}\right\}\nonumber \\
 && +\left.(b \leftrightarrow c, \, \nu \leftrightarrow \lambda, q \leftrightarrow r)+
	(b \leftrightarrow d, \nu \leftrightarrow \rho, q \leftrightarrow s) \right\} \nonumber \\
 && +\frac{g^2}{12}a^4\left\{\frac{2}{N_c}(\delta^{ab}\delta^{cd}+\delta^{ac}\delta^{bd}+\delta^{ad}
	\delta^{bc})\right.\nonumber \\
 && +\left.\sum_e (d^{abe}d^{cde}+d^{ace}d^{bde}+d^{ade}d^{bce})\right\}\nonumber \\
 &\times& \left\{\delta_{\mu\nu}\delta_{\mu\lambda}\delta_{\mu\rho}\xi^2_{\mu}
	\sum_{\sigma}\xi^2_{\sigma}\hat{k}_{\sigma}\hat{q}_{\sigma}\hat{r}_{\sigma}\hat{s}_{\sigma}-
	\delta_{\mu\nu}\delta_{\mu\lambda}\xi^2_{\mu}\xi^2_{\rho}\hat{k}_{\rho}\hat{q}_{\rho}\hat{r}_{\rho}
	\hat{s}_{\mu}\right. \nonumber \\
 && -\delta_{\mu\nu}\delta_{\mu\rho}\xi^2_{\mu}\xi^2_{\lambda}\hat{k}_{\lambda}\hat{q}_{\lambda}
	\hat{s}_{\lambda}\hat{r}_{\mu} 
  	-\delta_{\mu\lambda}\delta_{\mu\rho}\xi^2_{\mu}\xi^2_{\nu}\hat{k}_{\nu}\hat{r}_{\nu}\hat{s}_{\nu}
	\hat{q}_{\mu}-\delta_{\nu\lambda}\delta_{\nu\rho}\xi^2_{\mu}\xi^2_{\nu}\hat{q}_{\mu}\hat{r}_{\mu}
	\hat{s}_{\mu}\hat{k}_{\nu} \nonumber \\
 && +\delta_{\mu\nu}\delta_{\lambda\rho}\xi^2_{\mu}\xi^2_{\lambda}\hat{k}_{\lambda}\hat{q}_{\lambda}
	\hat{r}_{\mu}\hat{s}_{\mu}
  	+\left.\delta_{\mu\lambda}\delta_{\nu\rho}\xi^2_{\mu}\xi^2_{\nu}
	\hat{k}_{\nu}\hat{r}_{\nu}\hat{q}_{\mu}\hat{s}_{\mu}+\delta_{\mu\rho}\delta_{\nu\lambda}
	\xi^2_{\mu}\xi^2_{\nu}\hat{k}_{\nu}\hat{s}_{\nu}\hat{q}_{\mu}\hat{r}_{\mu}\right\} \nonumber 
\end{eqnarray}
\section{Bosonic Integral Evaluation}
\label{app:Integrals}
Apart from the denominators in 
Eq.~(\ref{g_pr}) and~(\ref{gh}), all the $\xi$ dependence in the evaluation of Feynman diagrams is 
polynomial in $\xi$ and $1/\xi$.
Having changed integration variables to the asymmetric 
Brillouin zone, only lattice zero-momentum integrals of the form
\be
	\B(q,n_1,n_2,n_3,n_4) = 
			\int_{-\pi}^\pi {d^4 k\over (2\pi)^4} {
			\hat{k}^{2n_1}_x \hat{k}^{2n_2}_y \hat{k}^{2n_3}_z \hat{k}^{2n_4}_t \over
			D_B(k,m)^q } 
	\label{Bint}
\ee
remain. 
In Eq.~(\ref{Bint}), $q$ and $n_i$ are positive integers, $\hat{k}_\mu = 2\sin(k_\mu/2)$ and the 
$\xi$ dependence is in the inverse bosonic propagator
\be
	D_B(k,m) = \sum_{\mu\in f} \hat{k}_\mu^2 +\frac{1}{\xi^2}\sum_{\mu\in c} 
	\hat{k}_\mu^2 + m^2 .
	\label{bosonicprop}
\ee
alone.
To evaluate these integrals, we adapt a technique given 
in Refs.~\onlinecite{Caracciolo:1992cp,Burgio:1996ji}. 
In the following, when one of the $n_i$ is zero it is
omitted as an argument of $\B$, while, where no confusion can arise, the index in $n_\mu$ is dropped 
and $\B^{f}$, $\B^{c}$ or $\B^{fc}$ denotes integrals whose numerators have a fine, coarse or mixed
momentum. 
In Refs.~\onlinecite{Caracciolo:1992cp,Burgio:1996ji}, a set of recursion relations
was defined to reduce every relevant integral to a linear combination of three
basic integrals. Although similar relations exist in the asymmetric case, the
lack of a complete symmetry among the indices, $n_i$ makes this reduction
more difficult. The relations can still be used to reduce the 
number of integrals which must be calculated considerably and to prove the cancellations of Lorentz-breaking 
terms. A first set can be obtained by expanding the trivial identity
\be
	\B(q-n,n_a,\ldots,n_b)=\int_{-\pi}^\pi {d^4 k\over (2 \pi)^4} {{\hat{k}^{2 n_a}_a \cdots 
				\hat{k}^{2 n_b}_b D_B(k,m)^n}\over D_B(k,m)^q }, 
	\label{Bintrel}
\ee
which gives relations of the type
\be
	(4-d)\B^f(q,1)+\frac{d}{\xi^2} \B^c(q,1) = \B(q-1) - m^2 \B(q),
\label{rel1}
\ee
\begin{eqnarray}
	(4-d)\B^f(q,2) + (4-d)(3-d)\B^f(q,1,1)+
        \frac{2 d (4-d)}{\xi^2}\B^{fc}(q,1,1)&+&\frac{d}{\xi^4}\B^c(q,2)+
		      \nonumber \\
    \frac{d (d-1)}{\xi^4}\B^c(q,1,1) = \B(q-2) - 2 m^2 \B(q-1) &+& m^4 \B(q)
	\label{bosonicrec1}
\end{eqnarray}
and so on. Furthermore, when $r>1$, using the identity 
\be
	{({\hat k}^2_\mu )^r \over D_B(k,m)^{q}} = 
			4 {({\hat k}^2_\mu)^{r-1} \over D_B(k,m)^{q}} + 2 
			\xi^2_{\mu }{({\hat k}^2_\mu  )^{r-2} 
			\over q - 1} \sin k_\mu  {\partial \over \partial k_\mu } { 1 \over D_B(k,m)^{q-1}} ,
\ee
and integrating by parts, we obtain the relation
\begin{eqnarray}
	\B(q,\ldots,r_\mu ,\ldots) &=& \xi^2_{\mu}{{r-1}\over{q-1}} \B(q-1,\ldots,r_\mu -1,\ldots) -
	\xi^2_{\mu }{{4r-6}\over{q-1}} \B(q-1,\ldots,r_\mu -2,\ldots) \nonumber \\
	& & +\, 4\, \B(q,\ldots,r_\mu -1,\ldots) , 
	\label{rec2}
\end{eqnarray}
which depends on the index $\mu =f,c$. A final set of relations is found by using
the trivial fact that the numerator cannot have more than four different arguments, e.g.
\be
	\B(q,1,1,1,1)\, = (d-4)\B(q+1,2_f,1,1,1) + 
        \frac{d}{\xi^2}\B(q+1,1,1,1,2_c) + m^2\,\B(q+1,1,1,1,1) .
	\label{identitabosonica}
\ee
The 2+2 case has the bonus relation, $1/\xi \B = \xi {\cal B}_{\frac{1}{\xi}}$
which is relevant to the numerical evaluation of a range of integrals. 
A similar relation would map 3+1 integrals to 1+3. With the help of these
relations we can reduce the propagator calculation to four converging integrals,
$\B(1)$, $\B^c(1,1)$, $\B^c(2,1)$ and $\B^f(2,1,1)$, the last of which vanishes 
for 3+1, and one infrared diverging integral, $\B(2)$.
 
For numerical calculation, setting $n=\sum_i n_i$ and using the 
well-known Schwinger representation~\cite{Kawai:1981ja}, the integrals are rewritten as
\be
	\B(q,n_1,n_2,n_3,n_4) = {{(-1)^n}\over 2^{q-n} 
                \Gamma(q)}\int_0^\infty d\lambda \;
                \lambda^{q-1} e^{-m^2\lambda/2} \prod_{\mu=1}^4 
		\left[\frac{d^{n_\mu}}{d x^{n_\mu}}
		\exp^{-x}I_0(x)\right]_{x=\frac{\lambda}{d_\mu^2}},
	\label{bessel}
\ee
with 
\be
	\frac{d^{n}}{d x^{n}}\exp^{-x}I_0(x) = \frac{e^{-x}}{2^{n-1}} 
                      \sum_{k=0}^n (-1)^{n-k} M_3^{n,k} I_k(x) ,
\ee 
where $M_3^{n,k}$ are partition multinomial coefficients and 
$I_k(x)$ are modified Bessel functions of the first kind~\cite{Book:Gradshteyn,Book:Abramovitz}. 
Since we want to focus attention on the massless case ($m^2\to 0$) and keeping only the 
non-vanishing terms either divergent or finite, power-counting shows that the integrals, 
$\B(q,n_1,n_2,n_3,n_4)$ for $q-n \le 1$, are infrared finite. 
As a result their value can be directly calculated from Eq.~(\ref{bessel}) by setting $m^2=0$. 
Numerical evaluation to, say, 32 digits becomes trivial with the use of a numerical
integration package available with programs like Mathematica or Maple (see Table~\ref{finite}). 
Furthermore, simple dimensional arguments give 
$\B(q,n_f,n_c)=O({(\xi\log\xi)^{q-n_f}}/\xi)$ for $q > n_f$ and 
$\B(q,n_f,n_c)=O(\xi^{q-n_f})$ for $q \leq n_f$ for the 2+2 case while for 3+1,
$\B(q,n_f,n_c)=O(\xi^{q-n_f})$. 
From this, the asymptotic behaviour of the one-loop corrections can be 
easily obtained, by noting
$\lim_{\xi \to \infty} \B^c(2,1)/(\xi^2 \B(1)) = 1/6$ for 3+1.
\begin{table}
\begin{tabular}{c c c}
\hline
$\xi$ & $2+2$ & $3+1$ \\
\hline
1 & 0.15493339023106021408483720810745 & 0.15493339023106021408483720810745\\
2 & 0.27309284159872576605248340464151 & 0.39746855267384293273601515359921\\
3 & 0.34489265148380372504682107682350 & 0.63909144650207181551150685543306\\
4 & 0.39476479072982726208009053119503 & 0.87599824804148552213829648851546\\
5 & 0.43269826887073920594692758485023 & 1.10989753717092264667311285543088\\
6 & 0.46325286965085016997908881189910 & 1.34197554748773397230022959566745\\
7 & 0.48882308585244885903279712081054 & 1.57290464634371239724254250061935\\
8 & 0.51080767194929084936132635888190 & 1.80307278894183022754245621178021\\
9 & 0.53009088018361126653278475943989 & 2.03271449270324135251760022607330\\
10& 0.54726616387832592956739420406785 & 2.26197832054321458404301146488387\\
\hline
\end{tabular}
\caption{$\B(1)$ for $2+2$ and $3+1$ as a function of $\xi$ to $10^{-32}$ precision.}
\label{finite}
\end{table}

We can now described the computation of the relevant part of 
$\B(q,n_1,n_2,n_3,n_4)$, $q-n \ge 2$, which is infrared divergent. 
Using the asymptotic expansion for large $x$ of $I_\nu(x)$ 
\be
	I_\nu(x)\approx \frac{\exp^{x}}{\sqrt{2\pi x}} \sum_{k=0}^\infty 
			\frac{(-1)^k\Gamma(\nu+k+1/2)}{(2 x)^{k} k! \Gamma(\nu-k+1/2)} ,
\ee
the leading and sub-leading behaviour is easily determined. 
This can be reexpressed in terms of $\Gamma(i,m^2), \,i=0\ldots q-n-2$ whose integral
representation can be directly subtracted from the integrand, leaving an $m^2\rightarrow0$ 
converging integral. For example, in the case $n=0$, the constants 
$b_i$ are defined as the expansion of $I_0(x)^4$, which
are rational numbers multiplied by $\pi^{-2}$.
The divergent part of $\B(q)$ is then given by
\be
	\frac{\xi^d}{\Gamma(q)}\sum_{i=2}^{q-1} \frac{b_{i-2} \Gamma(q-i)}{2^i (m^2)^{q-i}}
	   - \frac{\xi^d b_{q-2}}{2^p \Gamma(q)} \log m^2 \;.
	\label{divergenzabosonico}
\ee
Following the literature, the finite contribution to 
\be
	\B(2)=\frac{\xi^d}{16 \pi^2}(-\log m^2 -\gamma_E + F_0(\xi))+O(m^2) 
\ee  
is calculated first and the finite part of the other integrals is defined up 
to it.
Using 
\be
	\Gamma(0,m^2) = -\log m^2-\gamma_E -\sum_{n=1}^{\infty}\frac{(-m^2)^n}{nn!}
\ee
and taking the $\lim_{m^2\rightarrow 0}$, yields
\begin{eqnarray*}
	F_0(\xi)&=&\frac{4\pi^2}{\xi^d}\int_0^2 d\lambda \lambda \exp^{-(4-d+\frac{d}{\xi^2})\lambda}
		   I_0^{4-d}(\lambda) I_0^d(\frac{\lambda}{\xi^2})+\\
	        & & + \int_2^\infty  d\lambda \left(\frac{4\pi^2}{\xi^d}\lambda
		    \exp^{-(4-d+\frac{d}{\xi^2})\lambda}I_0^{4-d}(\lambda) I_0^d(\frac{\lambda}{\xi^2})-
		    \frac{1}{\lambda}\right) .
\end{eqnarray*}
\begin{table}
\begin{tabular}{c c c}
\hline
$\xi$ & $2+2$ & $3+1$ \\
\hline
1 & 4.3692252338747587180021767477 & 4.3692252338747587180021767478077\\
2 & 3.1818862274504847285875097585 & 3.0013807239614706751354255882375\\
3 & 2.2336373613788360098511237787 & 2.0763302182432342962706829473432\\
4 & 1.5519382904759206782491885601 & 1.4343462876751234663331108882378\\
5 & 1.0353940883292892812839278453 & 0.9504439561174006940006958674003\\ 
6 & 0.6235592513202110800048370509 & 0.5632462682606760634612041417967\\
7 & 0.2823597812415587497126918746 & 0.2405570583083765791881447111856\\ 
8 &-0.0084789916018349072700020647 & -0.036178339243301342993891362519\\
9 &-0.2617758427992475365501484308 & -0.278533198254716148526058151986\\
10&-0.4860763765593393395653061782 & -0.494190516782956896533394268235\\
\hline
\end{tabular}
\caption{$F_0(\xi)$ for $2+2$ and $3+1$ as a function of $\xi$ to $10^{-28}$ precision.}
\label{divergent}
\end{table}
For example 
\begin{eqnarray}
	\B(3) &=& {\xi^d\over 32\pi^2 m^2} + {\xi^d\over 128\pi^2}\left(\log m^2 + 
		   \gamma_E - F_0(\xi)\right) + f\B(3) , \\
	\B^f(3,1) &=& {\xi^d\over 64\pi^2} \left(-\log m^2 - \gamma_E + F_0(\xi)\right) + f\B^f(3,1)\\
	\B^c(3,1) &=& {\xi^{d+2}\over 64\pi^2} \left(-\log m^2 - \gamma_E + F_0(\xi)\right) + f\B^c(3,1).
\end{eqnarray}
which using Eq.~\ref{bosonicrec1} and taking 
$\lim_{m^2 \rightarrow 0} m^2 \B(3)$, must satisfy
\be
	(4-d)f\B^f(3,1) + \frac{d}{\xi^2} f\B^c(3,1) + \frac{\xi^d}{32 \pi^2} = 0 .
\ee
Numerical values for $\B(1)$ and $F_0$ are presented for a range of 
anisotropies in Tables~(\ref{finite},\ref{divergent}).

\begin{acknowledgments}
This work was funded by Enterprise-Ireland grants SC/2001/306 and
SC/2001/307. We would like to thank H. Panagopoulos for helpful discussions.
\end{acknowledgments}

\bibliography{main.bib}

\end{fmffile}
\end{document}